\documentclass[aps,prc,letterpaper,11pt,twoside,tightenlines,nofootinbib,showpacs,preprint]{revtex4}
\usepackage{graphicx}
\usepackage[sort&compress]{natbib}
\begin{document}

\newcommand{\be}{\begin{equation}}
\newcommand{\ee}{\end{equation}}
\newcommand{\bea}{\begin{eqnarray}}
\newcommand{\eea}{\end{eqnarray}}
\newcommand{\bef}{\begin{figure}}
\newcommand{\eef}{\end{figure}}
\newcommand{\bce}{\begin{center}}
\newcommand{\ece}{\end{center}}

\title{Energy weighted sum rules for mesons in hot and dense matter}

\author{D. Cabrera$^1$, A. Polls$^2$, A. Ramos$^2$ and L. Tol\'os $^3$\\
$^1$Departamento de F\'{\i}sica Te\'orica II, Universidad Complutense,\\
28040 Madrid, Spain\\
$^2$ Departament d'Estructura i Constituents de la Mat\`eria,\\
Universitat de Barcelona,
Diagonal 647, 08028 Barcelona, Spain \\
$^3$ Theory Group. KVI. University of Groningen, \\
Zernikelaan 25, 9747 AA Groningen, The Netherlands}

\begin{abstract}
We study energy weighted sum rules of the pion and kaon propagator
in nuclear matter at finite temperature. The sum rules are
obtained from matching the Dyson form of the
meson propagator with its spectral Lehmann representation at low and high 
energies. We calculate the sum
rules for specific models of the kaon and pion self-energy.
The in-medium spectral densities of the $K$ and $\bar K$ mesons 
are obtained
from a chiral unitary approach in coupled channels which incorporates the $S$-
and $P$-waves of the kaon-nucleon interaction. The pion self-energy is
determined from the $P$-wave coupling to particle-hole and $\Delta$-hole 
excitations, modified by
short range correlations. The sum rules for the lower energy weights are fulfilled
satisfactorily and reflect the contributions from the different 
quasi-particle and collective modes
of the meson spectral function. We discuss the sensitivity of the
sum rules to the distribution of spectral strength and their 
usefulness as quality tests of model calculations.
\end{abstract}
\vspace{1cm}

\date{\today}

\maketitle

\section{Introduction}
\label{sec:intro}

The properties of hadrons, both mesons and baryons, in hot and dense matter 
have been a matter of intense investigations over the last years
\cite{Rapp:1999ej,Fuchs:2005zg,Friedman:2007zz,Hayano:2008vn} and it is a
subject which is  calling the attention of many present and  future
experimental programs
\cite{exp}.

Of particular relevance are the lightest strange and non-strange mesons,
namely, kaons and pions. They typically appear as final state interacting
particles in nuclear production reactions. Being light, these pseudoscalar
mesons are also abundantly produced as thermal excitations in heavy-ion
collisions. Moreover, they constitute a relevant ingredient in the medium
modification of vector ($\rho$, $\omega$, $\phi$) and axial-vector ($a_1$)
mesons, as these decay strongly into light mesons and thus their in-medium
properties are tied to the modifications of the meson-cloud component of their
spectra. Vector mesons provide a unique tool to study high density and/or
temperature regions from electromagnetic decays, whereas the combined study of
the vector and axial-vector spectral functions  can shed light on the onset and
physical realization of chiral symmetry restoration in hot/dense strongly
interacting matter. Therefore, a solid knowledge of their interactions with the
medium through their coupling to light pseudoscalar mesons is mandatory.
Another relevant role of the properties of light mesons in nuclear matter is
played in the study of mesic atoms and nuclei, where the observation of bound
states of mesons and their spectral properties can lead to a better
understanding of meson meson and meson baryon interactions at finite nuclear
density.

One of the aims of many theoretical studies is to describe the propagation of
hadrons in hot and dense matter. A natural way to face this problem is to build
the single particle Green's function of the hadron. In turn, the
latter requires the  knowledge of the hadron self-energy, which
describes the interactions of the  particle with the medium. Obviously, the
quality of the calculation of the self-energy  relies on having a good model
for the interaction and a robust many-body framework. The single particle
Green's function has well defined analytical properties that impose some
constraints on both the many-body formalism and the interaction model. To
exploit these analytical  properties it is convenient to introduce the Lehmann
representation, which gives the Green's function in terms of the
single-particle spectral functions. An excellent tool to analyze these 
constraints is provided by the energy-weighted sum rules (EWSR) of the
single-particle spectral functions.   

Energy weighted sum rules have been extensively and successfully used in the
literature, mainly to analyze the response function of many-body systems in
particular for nuclear matter \cite{lipa89}, quantum liquids \cite{glyde94}
and more recently in the context of cold atoms \cite{dalfovo99}.  The energy
weighted moments of the response to a given operator allow one to estimate
the low energy states excited by this operator, specially for
highly collective states which concentrate a substantial amount of strength. An
important advantage of the sum rules is that they can be most of the times 
directly calculated without knowing the response function, by just evaluating
the expectation value in the ground state of commutators involving the
excitation operator and the Hamiltonian \cite{lipa89}.

In the context of chiral-symmetry breaking in hadronic physics, one finds a
prominent example in the 
well-known set of EWSR's proposed by
Weinberg \cite{Weinberg:1967kj}, which has been extended to hot and dense matter systems
\cite{Kapusta:1993hq}. The first of these relations connects the integrated difference of
the vector and axial vector mesonic spectral functions (current-current
correlators) with the pion decay constant.
Together with suitable model calculations, the first Weinberg sum rule can shed
light on the degree and physical mechanism of chiral symmetry restoration at
finite nuclear densities \cite{Kim:1999pb,cabrera-jido-roca-rapp}.

In the case of single-particle Green's functions, EWSR's for the nucleon
spectral functions have been since long-time well established in the
literature \cite{ripka}. However, only recently, the progress in the numerical
calculation of the single-particle spectral functions in nuclear matter has
permitted, through a careful analysis of the EWSR's, to identify the effects of
nucleon-nucleon correlations in the distribution of the single-particle
strength, both at zero \cite{Polls:1994zz} and finite
temperature \cite{Frick:2004jp,Rios:2006mn}.

However, EWSR's have not been used much for the case of meson single-particle
properties. They have been obtained in Ref.~\cite{DuttMazumder:2002me} for
$\omega$ mesons coupling to particle-hole excitations in a dense medium within
the long wavelength approximation. 
In this
paper, we present a derivation of the EWSR for the single-particle spectral
functions associated to the propagation of mesons in a hot and dense medium,
and discuss the physical implications of the fulfillment of these sum rules in
connection with the underlying interaction models as well as with certain
aspects of the meson nuclear phenomenology. Our aim is not only to analyze
the consistency of the many-body formalism used to calculate the meson Green's
function but also to obtain useful insights on the validity of the
meson-nucleon interaction model.

In Sect.~\ref{sec:formalism} we discuss the derivation of the EWSR's for
mesons propagating in cold nuclear matter and particularize for the case of
kaons and pions. In the case of kaons, the particle ($K$) and antiparticle
(${\bar K}$) modes behave differently in the nuclear medium,  
while the isotriplet pions exhibit a
common behavior in symmetric nuclear matter, which allows for a simplification
in the sum rules expressions. The generalization of the sum rules to 
nuclear matter at
finite temperature is also provided in this section. In
Sect.~\ref{sec:models} we briefly summarize particular models of the kaon
and pion self-energies in the nuclear medium, which have been discussed
elsewhere. The kaon self-energy is built from the effective kaon nucleon
interaction in $S$- and $P$-waves described in a chiral unitary approach
\cite{Tolos:2006ny,Tolos:2008di}.
In the case of pions, we consider the standard $P$-wave coupling to $ph$ and
$\Delta h$ configurations modified by spin-isospin short range correlations
\cite{Oset:1989ey,Oset:1981ih}.
The resulting sum rules for kaons and pions
are discussed in Sect.~\ref{sec:results}, for various momenta, nuclear densities
and temperatures. We analyze the
contribution of the different collective modes by studying the saturation of
the sum rules as a function of the energy, and discuss useful insights that can
be drawn for the particular self-energy models. A summary of our main 
conclusions is presented in Sect.~\ref{sec:conclusions}, together
with some discussion on the application of the present
method to test the consistency of the many-body scheme or the
meson nuclear interaction
model in a hot symmetric nuclear medium, as well as in other scenarios which
will be explored in the near future.

\section{Derivation of energy weighted sum rules}
\label{sec:formalism}

\subsection{Zero temperature}
\label{ssec:zerot}

The derivation of the energy weighted sum rules (EWSRs) for hadrons and, in
particular, for mesons follows from comparing the in-medium
propagator with the corresponding Lehmann spectral representation. The propagator for a meson
$M$ of energy $q^0$ and momentum $\vec{q}\,$ in symmetric
nuclear matter of density $\rho$ reads:
\bea
D_{M } (q^0,\vec{q}\,;\rho) &=& \frac{1}{(q^0)^2 - 
\omega_{M}^2(\vec{q}\,) - \Pi_{M}(q^0,\vec{q}\,;\rho)} \ ,
\label{eq:prop_dyson}
\eea 
where $\omega_{M }(\vec{q}\,)=\sqrt{m^2+\vec{q}\,^2}$ is the free
dispersion relation and $\Pi_{M}$ the meson self-energy.
The corresponding spectral (Lehmann) representation when 
the meson and antimeson behave as distinct particles in the medium is
\bea
D_{M} (q^0,\vec{q}\,;\rho) &=& \int_{0}^{\infty}
\textrm{d}\omega \, 
\left\{
\frac{S_{M}(\omega,\vec{q}\,;\rho)}{q^0 - \omega + {\rm i}\eta}
-
\frac{S_{\bar M}(\omega,\vec{q}\,;\rho)}{q^0 + \omega + {\rm i}\eta}
\right\} \ ,
\label{eq:prop_lehmann}
\eea
with 
\bea
S_{M}(q^0,\vec{q}\,;\rho)=-\frac{1}{\pi} {\rm Im}\, D_{M}
(q^0,\vec{q}\,;\rho) \ .
\eea

We start by analyzing the $q^0\to\infty$ limit. In order to obtain 
the
expansion of the propagator in powers of $1/q^0$ we first study the
behavior of the self-energy $\Pi_{M}(q^0,\vec{q}\,;\rho)$ at high
energies from its dispersion relation
\bea  
\Pi_{M}(q^0,\vec{q}\,;\rho)= \Pi_{M}^{\infty}(\vec{q}\,;\rho)
- \frac{1}{\pi}\int_{-\infty}^{\infty}
\textrm{d}\omega \frac{{\rm Im}\,\Pi_{M}(\omega,\vec{q}\,;\rho)}{q^0-\omega+i\eta} \ ,
\label{eq:pi_dispersion}
\eea
where $\Pi_{M}^{\infty}$ is the real non-dispersive contribution of the self-energy.
In the particular models discussed in the next section,
this quantity will be either zero or stay finite.
By expanding the real part of Eq.~(\ref{eq:pi_dispersion})
around $q^0\to \infty$ we obtain:
\bea 
{\rm Re}\,\Pi_{M}(q^0,\vec{q}\,;\rho) &=& \Pi_{M}^{\infty}(\vec{q}\,;\rho)
-\frac{1}{\pi} \frac{1}{q^0} \left[\int_{-\infty}^\infty \textrm{d}\omega\, 
{\rm Im}\,\Pi_{M}(\omega,\vec{q}\,;\rho) \right. \nonumber  \\ 
&+& \left. \frac{1}{q^0} \int_{-\infty}^\infty \textrm{d}\omega \,
\omega \, {\rm Im}\,\Pi_{M}(\omega,\vec{q}\,;\rho) +
\frac{1}{(q^0)^2} \int_{-\infty}^\infty \textrm{d}\omega \, 
\omega^2 \, {\rm Im}\,\Pi_{M}(\omega,\vec{q}\,;\rho) + \dots \right]
\ .
\label{eq:pi_expansion}
\eea
Using the properties of the retarded self-energy:
\bea
{\rm Re}\,\Pi_{M}(-q^0,\vec{q}\,;\rho)&=& {\rm Re}\,\Pi_{\bar
M}(q^0,\vec{q}\,;\rho) \nonumber \\
{\rm Im}\,\Pi_{M}(-q^0,\vec{q}\,;\rho)&=& -{\rm Im}\,\Pi_{\bar
M}(q^0,\vec{q}\,;\rho) \ ,
\label{eq:crossing}
\eea
we can rewrite Eq.~(\ref{eq:pi_expansion}) as
\bea 
{\rm Re}\,\Pi_{M}(q^0,\vec{q}\,;\rho) &=& \Pi_{M}^{\infty}(\vec{q}\,;\rho)
-\frac{1}{\pi} \frac{1}{q^0}   \left\{ \int_0^\infty \textrm{d}\omega \,
[{\rm Im}\,\Pi_{M }(\omega,\vec{q}\,;\rho)
-{\rm Im}\,\Pi_{\bar M}(\omega,\vec{q}\,;\rho)] \right. \nonumber \\ 
& + & \frac{1}{q^0} \int_0^\infty \textrm{d}\omega \, \omega \,  
[{\rm Im}\,\Pi_{M}(\omega,\vec{q}\,;\rho)+
{\rm Im}\,\Pi_{\bar M}(\omega,\vec{q}\,;\rho)] \nonumber \\
& + &
 \left.  \frac{1}{(q^0)^2} \int_0^\infty \textrm{d}\omega \, \omega^2 \, 
[{\rm Im}\,\Pi_{M}(\omega,\vec{q}\,;\rho) -
{\rm Im}\,\Pi_{\bar M}(\omega,\vec{q}\,;\rho)]
 + \dots \right\} \ .
\label{eq:pi_expansion2}
\eea
Accordingly, the first few terms of the expansion of the real part 
of the in-medium propagator [Eq.~(\ref{eq:prop_dyson})] read:
\bea 
{\rm Re}\,D_{M}(q^0,\vec{q}\,;\rho) =
\frac{1}{(q^0)^2}\left\{1 \right. &+ & \frac{1}{(q^0)^2}
[\omega_{M}^2(\vec{q}\,)+
\Pi_{M}^{\infty}(\vec{q}\,;\rho)] \nonumber \\
 &-& 
\frac{1}{\pi} \frac{1}{(q^0)^3} \int_0^\infty \textrm{d}\omega \,
[{\rm Im}\,\Pi_{M }(\omega,\vec{q}\,;\rho)
-{\rm Im}\,\Pi_{\bar M}(\omega,\vec{q}\,;\rho)] \nonumber \\
&+& 
\frac{1}{(q^0)^4}\left([\omega_{M}^2(\vec{q}\,)+
\Pi_{M}^{\infty}(\vec{q}\,;\rho)]^2 \right. \nonumber \\
&& ~~~~~~~~  \left. \left. - \frac{1}{\pi}\int_0^\infty 
\textrm{d}\omega \, \omega \, 
[{\rm Im}\,\Pi_{M}(\omega,\vec{q}\,;\rho)+
{\rm Im}\,\Pi_{\bar M}(\omega,\vec{q}\,;\rho)]  \right) + \dots \right\}\ .
\nonumber \\
\label{eq:dyson_expansion}
\eea
On the other hand, we obtain the following
expansion around $q^0 \to \infty$ from the Lehmann representation
[Eq.~(\ref{eq:prop_lehmann})]:
\bea
\label{eq:lehmann-exp}
{\rm Re}\,D_{M}(q^0,\vec{q}\,;\rho)
&=&
\frac{1}{q^0} \, \sum_{n=0}^{\infty} \, \int_0^{\infty}
\textrm{d}\omega \, \left[ \frac{\omega}{q^0} \right]^{2n}
\, [ S_{M}(\omega,\vec{q}\,;\rho) - S_{\bar M}(\omega,\vec{q}\,;\rho) ]
\nonumber \\
&+&
\frac{1}{q^0} \, \sum_{m=0}^{\infty} \, \int_0^{\infty}
\textrm{d}\omega \, \left[ \frac{\omega}{q^0} \right]^{2m+1}
\, [ S_{M}(\omega,\vec{q}\,;\rho) + S_{\bar M}(\omega,\vec{q}\,;\rho) ]
\ ,
\eea
which displays separately the terms involving the sum and the difference of 
the meson and antimeson
spectral functions. 
The sum rules are readily obtained from matching
Eqs.~(\ref{eq:dyson_expansion}) and
(\ref{eq:lehmann-exp}), order by
order in $1/q^0$. The first few terms up to $(1/q^0)^4$ determine:
\bea
m_{0}^{(\mp)}(q;\rho)&:&  \ \ (n=0)\ \ 
\int_0^{\infty} \textrm{d}\omega \,
[ S_{M}(\omega,\vec{q}\,;\rho) - S_{\bar M}(\omega,\vec{q}\,;\rho) ] = 0
 \label{eq:sum_zero_m}
 \\
& & \ \ (m=0)\ \ 
\int_0^{\infty} \textrm{d}\omega \, \omega \, [ S_{M}(\omega,\vec{q}\,;\rho) 
+ S_{\bar M}(\omega,\vec{q}\,;\rho) ] = 1 \ , 
\label{eq:sum_zero_p}
\\
m_{1}^{(\mp)}(q;\rho)&:&  \ \ (n=1)\ \ 
\int_0^{\infty} \textrm{d}\omega \,
\omega^2 \, [ S_{M}(\omega,\vec{q}\,;\rho) - S_{\bar M}(\omega,\vec{q}\,;\rho)]
= 0
 \label{eq:sum_one_m}
 \\
& & \ \ (m=1)\ \ 
\int_0^{\infty} \textrm{d}\omega \,
\omega^3 \, [ S_{M}(\omega,\vec{q}\,;\rho) + S_{\bar M}(\omega,\vec{q}\,;\rho)
] =
\omega_{M}^2(\vec{q}\,) + \Pi_{M}^{\infty}(\vec{q}\,;\rho) \ .
\label{eq:sum_one_p}
\eea
The dispersive part of the self-energy contributes to the right hand side starting from
$(1/q^0)^5$. For instance, in the case $n=m=2$ the sum rules read:
\bea
&& m_{2}^{(\mp)}(q;\rho): \nonumber \\
&& (n=2)\ \ 
\int_0^{\infty} \textrm{d}\omega \,
\omega^4 \, [ S_{M}(\omega,\vec{q}\,;\rho) - S_{\bar M}(\omega,\vec{q}\,;\rho) ]
= -\frac{1}{\pi} \int_0^\infty \textrm{d}\omega \,
[{\rm Im}\,\Pi_{M}(\omega,\vec{q}\,;\rho)
-{\rm Im}\,\Pi_{\bar M}(\omega,\vec{q}\,;\rho)]
 \nonumber \\
 \\
 \label{eq:sum_two_m}
&& (m=2)\ \  
\int_0^{\infty} \textrm{d}\omega \,
\omega^5 \, [ S_{M}(\omega,\vec{q}\,;\rho) + S_{\bar M}(\omega,\vec{q}\,;\rho)
] =
[\,\omega_{M}^2(\vec{q}\,)+
\Pi_{M}^{\infty}(\vec{q}\,;\rho)]^2 \nonumber \\
&&
\phantom{ (m=2)\ \  
\int_0^{\infty} \textrm{d}\omega \,
\omega^5 \, [ S_{M}(\omega,\vec{q}\,;\rho) +
] =}
- \frac{1}{\pi}\int_0^\infty \textrm{d}\omega \, \omega  \,
[{\rm Im}\,\Pi_{M}(\omega,\vec{q}\,;\rho)+
{\rm Im}\,\Pi_{\bar M}(\omega,\vec{q}\,;\rho)] \ .
\label{eq:sum_two_p}
\eea
Furthermore, another sum rule results from the evaluation of the zero-mode
propagator, namely $q^0=0$:
\bea
m_{-1}(q;\rho)&:&  \ \ \ 
\int_0^{\infty} \textrm{d}\omega  \, \frac{1}{\omega}\,
[S_{M}(\omega,\vec{q}\,;\rho) + S_{\bar M}(\omega,\vec{q}\,;\rho)] 
= \frac{1}{\omega_{M}^2(\vec{q}\,) + \Pi_{M}(0,\vec{q}\,;\rho)}
\ .
\label{eq:sum_minusone}
\eea
Note that, as implied by the sum rules $m_{-1}$ and $m_{1}^{(+)}$,
the self-energy $\Pi_{M}(q^0,\vec{q}\,;\rho)$ at $q^0=0$ and $q^0\to
\infty$ is necessarily real. One expects
this from the phenomenological point of view: at zero energy there should not be
any open in-medium channel for the meson to decay into, whereas at high energies 
form-factors or cut-offs are usually applied to truncate
the modes not accounted for explicitly by the hadronic model. 
Also note that the sum rules have to be satisfied for every value of the meson
momentum, $q$. The $m_0$ sum rule is of particular relevance, since it 
is a consequence of the canonical commutation relation of the meson field \cite{walecka}.

The sum rules given by Eqs. (\ref{eq:sum_zero_m}) to (\ref{eq:sum_minusone})
are valid for the general case in which the meson and the antimeson
behave differently in the nuclear medium, such as kaons and antikaons, or pions
in asymmetric nuclear matter. For
instance, by substituting $M \to \bar K$ and $\bar M \to K$ the former expressions 
would represent the sum rules for the antikaon. Similar expressions would be
obtained for the kaon case if $M \to K$ and $\bar M \to \bar K$. 
It is worth noticing that, even if the $\bar K N$ and
$KN$ interactions are very different in nuclear matter, the EWSRs impose
constraints on the behavior of the $\bar K$ and $K$ self-energies.
In particular and due to the symmetry under the exchange $K \leftrightarrow {\bar
K}$ on the l.h.s., the $m_{-1}$ and $m_{1}^{(+)}$ sum rules indicate that
not only $\Pi_{\bar K}(q^0,\vec{q}\,)$ and  $\Pi_{K}(q^0,\vec{q}\,)$ should be
real for $q^0=0$ and $q^0 \rightarrow \infty$, but also that both must 
coincide at the corresponding low- and high-energy limits. Actually, this is a
consequence of the crossing symmetry relations given in Eq.~(\ref{eq:crossing}).
Therefore, sum rules obtained from models that do not respect 
this symmetry will only be fulfilled to a certain level, depending on the severity of the 
violation. 

In the particular case of pions in symmetric matter, we have
$\Pi_{M}(q^0,\vec{q}\,;\rho)=\Pi_{\bar M}(q^0,\vec{q}\,;\rho)$
since particles and
antiparticles behave identically. Consequently, 
only the even powers of $1/q^0$  survive in the expansion of the propagator and, 
correspondingly, the sum rules acquire the following simplified forms:
\bea
m_{0}(q ; \rho)&:&  \ \ \ 
\int_0^{\infty} \textrm{d}\omega \,
2\omega \, S_{\pi}(\omega,\vec{q}\,;\rho) = 1 \ ,
\\ 
m_{1}(q; \rho)&:&  \ \ \ 
\int_0^{\infty} \textrm{d}\omega \,
2\omega^3 \, S_{\pi}(\omega,\vec{q}\,;\rho) 
=
\omega_{\pi}^2(\vec{q}\,) + \Pi_{\pi}^{\infty}(\vec{q}\,;\rho) \ ,
\\
m_{-1}(q;\rho)&:& \ \ \ 
\int_0^{\infty} \textrm{d}\omega \,
\frac{2}{\omega} \, S_{\pi}(\omega,\vec{q}\,;\rho)
=
\frac{1}{\omega_{\pi}^2(\vec{q}\,) + \Pi_{\pi}(0,\vec{q}\,;\rho)}.
\eea

\subsection{Finite temperature}
\label{ssec:finitet}

The extension of the EWSRs to finite temperature $T$ is straightforward. We
elaborate on this for the antikaon case below, whereas for the pion case
the derivation is completely similar.
Once again, the sum rules are obtained from the expansion at high energy
of both the Dyson form of the propagator and its Lehmann
representation. At finite $T$, the spectral representation
is obtained in the Matsubara space, namely
\bea
D_{\bar K}(\omega_n,\vec{q}\,;\rho,T)=-\frac{1}{\pi} \int^{\infty}_{-\infty} 
\textrm{d}\omega \,
\frac{{\rm Im} \, D_{\bar K}(\omega,\vec{q};\rho,T)}{{\rm i} \omega_n-\omega} \
,
\eea
where ${\rm i} \omega_n=2n\pi T$ is the bosonic Matsubara frequency.
If we now split the integral in two pieces,
\bea
D_{\bar K}(\omega_n,\vec{q}\,;\rho,T)=-\frac{1}{\pi} \left[
\int^{0}_{-\infty} \textrm{d}\omega \,
\frac{{\rm Im}\, D_{\bar K}(\omega,\vec{q}\,;\rho,T)}{{\rm i} \omega_n-\omega} +
\int^{\infty}_{0}  \textrm{d}\omega \,
\frac{{\rm Im}\, D_{\bar K}(\omega,\vec{q}\,;\rho,T)}{{\rm i} \omega_n-\omega} \right] \ ,
\eea
and change $\omega \rightarrow -\omega$ in the first term, we obtain
\bea
D_{\bar K}(\omega_n,\vec{q}\,;\rho,T)=-\frac{1}{\pi} \left[
\int^{\infty}_{0} \textrm{d}\omega \,
\frac{{\rm Im}\, D_{\bar K}(-\omega,\vec{q}\,;\rho,T)}{{\rm i} \omega_n+\omega} +
\int^{\infty}_{0}  \textrm{d}\omega \,
\frac{{\rm Im}\, D_{\bar K}(\omega,\vec{q}\,;\rho,T)}{{\rm i} \omega_n-\omega} \right] \ .
\eea
In a fully relativistic thermal calculation, the imaginary part of the retarded
self-energy satisfies
\bea
{\rm Im} \, \Pi_{\bar K} (-\omega, \vec{q}\,;\rho,T)
=
-{\rm Im} \, \Pi_{ K} (\omega, \vec{q}\,;\rho,T) \ ,
\eea
and, hence, the same applies to the spectral function.
Thus, the spectral representation of the propagator actually reads
\bea
D_{\bar K}(\omega_n,\vec{q}\,;\rho,T)=-\frac{1}{\pi} \left[
\int^{\infty}_{0} \textrm{d}\omega \,
\frac{{\rm Im} \, D_{\bar K}(\omega,\vec{q};\rho,T)}{{\rm i} \omega_n-\omega} -
\int^{\infty}_{0}  \textrm{d}\omega \,
\frac{{\rm Im} \, D_{K}(\omega,\vec{q}\,;\rho,T)}{{\rm i} \omega_n+\omega} \right] \ .
\eea
We then perform the analytical continuation onto the real axis, ${\rm
i}\omega_n \rightarrow q^0+{\rm i}\eta$, and one finally gets
\bea
D_{\bar K} (q^0,\vec{q}\,;\rho,T) &=& \int_{0}^{\infty}
\textrm{d}\omega \, 
\left\{
\frac{S_{\bar K}(\omega,\vec{q}\,;\rho,T)}{q^0- \omega+{\rm i}\eta }
-
\frac{S_{K}(\omega,\vec{q}\,;\rho,T)}{q^0 + \omega+{\rm i}\eta}
\right\} \ ,
\label{dft}
\eea
where
\bea
S_{\bar K (K)}(\omega,\vec{q}\,;\rho,T)=-\frac{1}{\pi} {\rm Im} \, D_{\bar K (K)} (\omega,\vec{q}\,;\rho,T) \ .
\eea
The expression of Eq.~(\ref{dft}) for the spectral representation of the
propagator has the same behavior at $q^0=0$ and $q^0\to\infty$ as the one
obtained in the $T=0$ case.
Therefore, the EWSRs at finite density and temperature have the
same form as the ones at $T=0$. Summarizing, for $m_{-1}$, $m_0^{(\mp)}$ and
$m_1^{(\mp)}$ one has

 \bea
m_{-1}(q;\rho,T)&:& \ \ \ 
\int_0^{\infty} \textrm{d}\omega \,
\frac{1}{\omega} \, [ S_{\bar K}(\omega,\vec{q}\,;\rho,T) + S_{K}(\omega,\vec{q}\,;\rho,T)]
=
\frac{1}{\omega_{K}^2(\vec{q}\,) + \Pi_{\bar K}(0,\vec{q}\,;\rho,T)} \ ,
\nonumber \\
\\
m_{0}^{(\mp)}(q;\rho,T)&:&  \ \ \ 
\int_0^{\infty} \textrm{d}\omega \,
[ S_{\bar K}(\omega,\vec{q}\;\rho,T) - S_{K}(\omega,\vec{q}\,;\rho,T) ] = 0
\nonumber \\
& & \ \ \ 
\int_0^{\infty} \textrm{d}\omega \,
\omega \, [ S_{\bar K}(\omega,\vec{q}\,;\rho,T) + S_{K}(\omega,\vec{q}\,;\rho, T) ] = 1 \ ,
\nonumber \\
\\
m_{1}^{(\mp)}(q;\rho,T)&:&  \ \ \ 
\int_0^{\infty} \textrm{d}\omega \,
\omega^2 \, [ S_{\bar K}(\omega,\vec{q}\,;\rho,T) - S_{K}(\omega,\vec{q}\,;\rho,T) ] = 0
\nonumber \\
& & \ \ \ 
\int_0^{\infty} \textrm{d}\omega \,
\omega^3 \, [ S_{\bar K}(\omega,\vec{q}\,;\rho,T) + S_{K}(\omega,\vec{q}\,;\rho,T) ] 
=
\omega_{K}^2(\vec{q}\,) + \Pi_{\bar K}^{\infty}(\vec{q}\,;\rho,T) \ .
\nonumber \\
\eea

\section{\mbox{\boldmath{$\bar K$, $K$}} and pion self-energy models}
\label{sec:models}

The EWSRs constitute an ideal test of the quality of any hadronic model. The
energy weighted integrals of the hadronic spectral function, on the one side,
are compared to the low and high energy limits of the corresponding self-energy
or to model-independent values, on the other side.

Here we briefly recall the essential features of recent calculations of the
properties of kaons in dense matter at zero and finite temperature.
We refer to
Refs.~\cite{Tolos:2006ny,Tolos:2008di} for details.
The $\bar K$ and $K$ self-energies in symmetric nuclear matter at finite
temperature are obtained from an evaluation of the in-medium kaon-nucleon
interaction within a chiral unitary approach. The model incorporates the 
$S$- and
$P$-waves of the kaon-nucleon interaction. 

At tree level, the $S$-wave
amplitude arises from the Weinberg-Tomozawa term of the chiral Lagrangian.
Unitarization in coupled channels is imposed by solving the Bethe-Salpeter
equation with on-shell amplitudes. With a single regularization parameter, the
unitarized $\bar K N$ amplitude generates dynamically the $\Lambda(1405)$
resonance in the $I=0$ channel and provides a satisfactory description of
low-energy scattering observables.
The in-medium solution of the $S$-wave amplitude accounts for Pauli-blocking
effects, mean-field binding on the nucleons and hyperons via a $\sigma-\omega$
model, and the dressing of the pion and kaon propagators through their corresponding
self-energies, in a self-consistent manner.
The relation
\begin{eqnarray}
\Pi^s_{\bar K(K)}(q_0,{\vec q};T)= \int \frac{d^3p}{(2\pi)^3}\,
n_N(\vec{p},T) \, [{T}^{(I=0)}_{\bar K(K)N}(P_0,\vec{P};T) +
3{T}^{(I=1)}_{\bar K(K)N}(P_0,\vec{P};T)]\ . \label{eq:selfd}
\end{eqnarray}
determines the
antikaon (kaon) dominant $S$-wave component of the self-energy in terms of 
the in-medium effective
antikaon(kaon)-nucleon interaction in $S$-wave.

We should mention that the loop
integrals are regularized by a cut-off momentum of $q_{\rm max}=630$ MeV/c. 
This means that the right-hand side
unitary cut is correctly implemented up to center-of-mass energies $\sqrt{s}$ of
about 2 GeV, above which the model cannot be trusted. This in turn imposes a
limit of $q^0 \sim 1$ GeV for the calculated self-energies of the $K$ and $\bar
K$ mesons. We will make sure in the next section that energies 
beyond this range no longer contribute to the sum rule under study.

The model incorporates, in addition, a $P$-wave contribution to
the self-energy from
hyperon-hole ($Yh$) excitations, including $\Lambda$, $\Sigma$ and
$\Sigma^*$ components.

Finite temperature effects are implemented in the intermediate meson-baryon
states following the Imaginary Time Formalism, thus keeping the analytical constraints
of the retarded self-energies of the $K$ and $\bar K$ mesons.

The results from Ref.~\cite{Tolos:2008di} show that the $\bar K$ effective mass
gets lowered by about $50$~MeV in cold nuclear matter at saturation density,
whereas finite temperature reduces this attraction to 50\% at $T=100$~MeV. The
$P$-wave contribution to the ${\bar K}$ optical potential, due to $\Lambda$,
$\Sigma$ and $\Sigma^*$ excitations, becomes significant for momenta larger
than 200 MeV/c and softens the attraction felt by the $\bar K$ in the nuclear
medium moderately. The $\bar K$ spectral function spreads over a wide range of
energies, reflecting the melting of the $\Lambda (1405)$ resonance and the
$Yh$ contributions at finite temperature. Regarding the $K$
self-energy, it is found that the low-density theorem is a good approximation
close to saturation density, due to the absence of resonance-hole excitations
in the $KN$ interaction. The $K$ potential shows a moderate repulsive behavior,
whereas the quasi-particle peak is considerably broadened with increasing
density and temperature. Implications of these results for the decay of the
$\phi$ meson and transport calculations in Heavy-Ion Collisions were also
discussed in Ref.~\cite{Tolos:2008di}.

Next, we briefly discuss the many-body mechanisms included in the modification
of the pion propagator in a nuclear medium. In cold nuclear matter, the
pion spectral function exhibits a mixture of the pion quasi-particle mode and
particle-hole
($ph$), Delta-hole ($\Delta h$) excitations. Following the calculation in \cite{Oset:1989ey}
(extended to finite temperatures in \cite{Tolos:2008di}), the lowest order
irreducible $P$-wave pion self-energy reads 
\begin{equation}
\label{eq:piself-ph-Dh} \Pi_{\pi NN^{-1}+\pi\Delta N^{-1}}^p
(q_0,\vec{q};\rho,T) = \left( \frac{f_N}{m_{\pi}} \right) ^2 \vec{q}\,^2 \,
\left[ U_{NN^{-1}} (q_0,\vec{q};\rho,T) + U_{\Delta N^{-1}}
(q_0,\vec{q};\rho,T) \right] \,\,\, , 
\end{equation} 
where the finite
temperature Lindhard functions for the $ph$ and $\Delta h$ excitations are
given in detail in the appendix of Ref.~\cite{Tolos:2008di}.
The strength of the collective modes excited by the pion is further modified by
repulsive, spin-isospin $NN$ and $N\Delta$ short range
correlations \cite{Oset:1981ih}, which we include in a
phenomenological way with a Landau-Migdal effective interaction.

At normal nuclear matter density, the pion spectral function clearly exhibits
the different modes excited in the medium. At low momentum, the pion
quasi-particle peak carries most of the strength together with a moderate
contribution of the
$ph$ excitations at lower energies.
The pion mode feels a sizable attraction with respect to
that in free space. At larger momentum values of a few hundred MeV/c, 
the excitation of the
$\Delta h$ mechanism takes over and provides a considerable amount of
strength overlapping with the pion quasi-particle peak which 
broadens considerably.
 
At finite temperatures, the
softening of the nucleon occupation number due to thermal motion
causes a broadening of the three modes present in the spectral
function.
In the next section we discuss how these features of the pion spectral
function reflect in the saturation of the different EWSRs.

\section{Results and discussion}
\label{sec:results}

In this section we analyze the behavior of the energy weighted sum rules for
the particular models of the kaon and pion properties in a hot and dense nuclear
medium described in the former section. As commonly done
\cite{Polls:1994zz,Frick:2004jp,Rios:2006mn}, we depict the left-hand side
(l.h.s.) of each sum rule as a function of the upper limit of the energy
integration. This allows us to examine how relevant is the contribution of the
different modes populating the meson spectral function  (collective modes,
quasi-particle peak) in saturating the sum rule. Note that depending on the
energy weight of the sum rule different energy regions will be scanned.
The horizontal scale ends at 1000 MeV since, as noted in the previous section,
the model calculation of the $K({\bar K})$ self-energy cannot be trusted beyond
this energy value.

\begin{figure}[t]
\begin{center}
\includegraphics[width=0.9\textwidth]{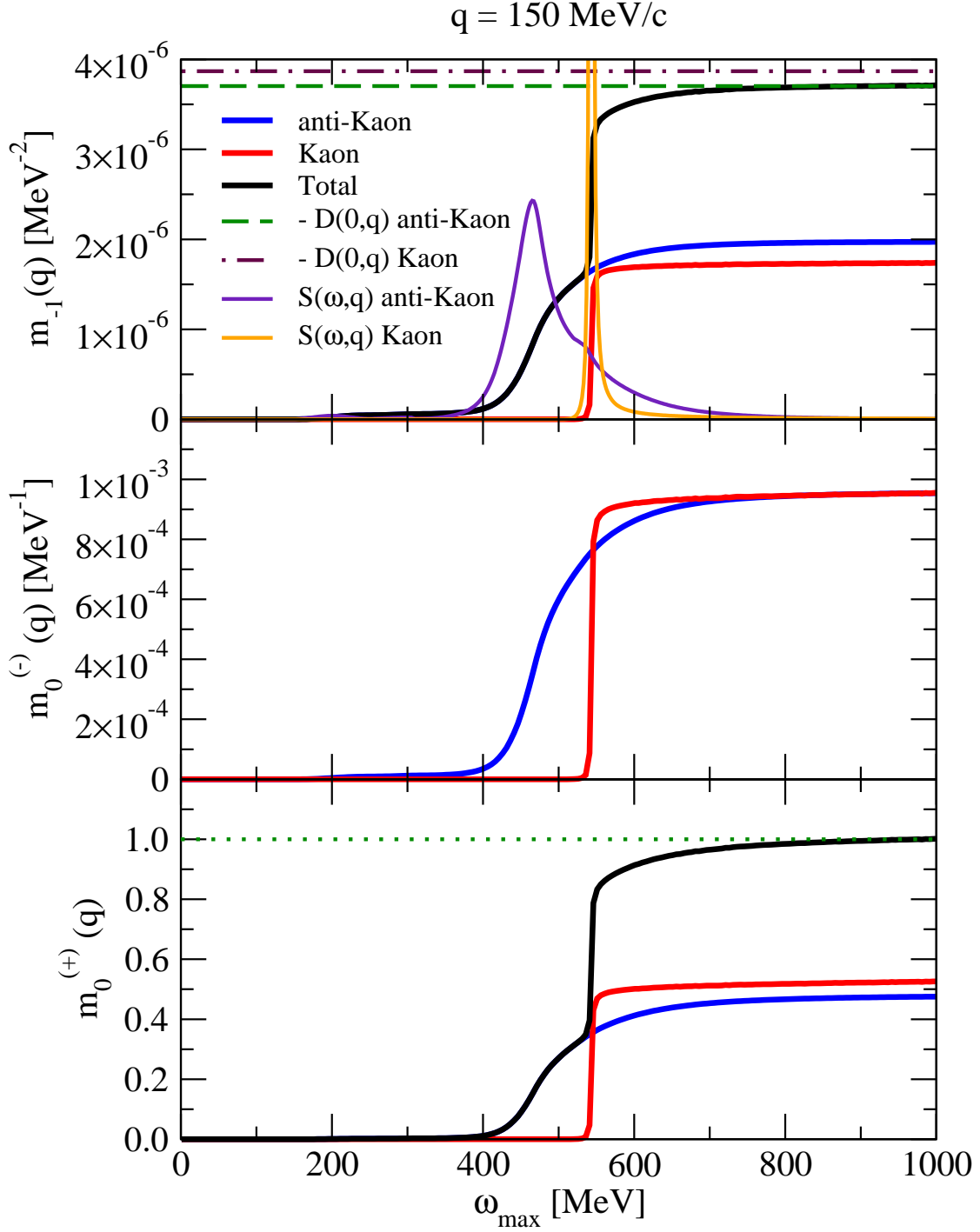}
\caption{(Color online) $m_{-1}$, $m_{0}^{(-)}$ and $m_{0}^{(+)}$ sum rules for the $K$ and $\bar K$ spectral functions at
$q=150$~MeV/c, $\rho=\rho_0$ and zero temperature. The $\bar K$, $K$ spectral
functions are also displayed for reference in arbitrary units.
Note that the $m_{0}^{(+)}$ sum rule (lower panel) is independent of the meson
momentum.}
\label{fig:kaon-cold}
\end{center}
\end{figure}

The 
$m_{-1}$, $m_0^{(-)}$ and $m_0^{(+)}$ sum rules for the antikaon propagator
are shown in Fig.~\ref{fig:kaon-cold} in the case
of normal nuclear matter density, zero temperature and 150~MeV/c kaon
momentum. The contributions from $\bar K$ and $K$ to the l.h.s. of the sum rule,
cf.~Eqs.~(\ref{eq:sum_zero_m}), (\ref{eq:sum_zero_p}), and
(\ref{eq:sum_minusone}), 
are depicted
separately. The $\bar K$ and $K$ spectral functions are also shown for
reference in arbitrary units.

The l.h.s. of the $m_{-1}$ sum rule (upper panel) converges properly and saturates a few
hundred MeV beyond the quasiparticle peak. The antikaon part has a soft
behavior as the $\bar K$ spectral function spreads as a consequence of the
mixing of the quasiparticle peak and the $\Lambda(1405) h$ mode. Note
that the subthreshold $P$-wave $Yh$ components, although small at low
momentum, have a visible contribution to the sum rule below the quasi-particle
peak, as a consequence of the $\omega^{-1}$ energy weight in the integrand of
Eq.~(\ref{eq:sum_minusone}). The $K$ contribution carries about
half the weight of the saturated sum rule, as it would be the case
in the absence of interactions. The contribution, which is mainly
concentrated at the quasi-particle energy, reflects the narrowness
of the $K$ spectral function even at normal matter density, as no baryonic
resonances in the $S=+1$ channel can be excited. 

We have also plotted in
Fig.~\ref{fig:kaon-cold} the right hand side (r.h.s.) of the $m_{-1}$ sum
rule both for the antikaon and kaon, namely their off-shell propagators
evaluated at zero energy (modulo a minus sign).
The difference between both values indicates $\Pi_K(q^0=0,\vec{q};\rho) \neq \Pi_{\bar
K}(q^0=0,\vec{q};\rho)$, which reflects the violation of
crossing symmetry present in the chiral model employed
for the kaon and antikaon
self-energies. Although this model works well in the time-like region for 
kaon (antikaon) energies from $m_K$ to about 1~GeV, its limitations show up 
for space-like kaons (antikaons) 
since their zero-mode propagators do not coincide.

We recall that, in fact, the chiral $K (\bar K)N$ amplitudes are dominated by the
unitarized $S$-wave component which is built by neglecting the explicit
exchange of a meson-baryon pair in a $t$-channel configuration, thus violating
crossing symmetry. This approximation plays a minor role in the $S$-wave
amplitudes at energies around the ${K} (\bar K) N$ threshold, but may turn
relevant for the largely off-shell amplitudes explored in the evaluation of the
$K(\bar K)$ propagator at $q^0=0$ MeV. Having this in mind, we may still expect
the saturated value of the l.h.s. of the $m_{-1}$ sum-rule to provide a
constraint for the value of the zero-mode propagator appearing on the r.h.s.
This is so because, as seen in Fig.~\ref{fig:kaon-cold}, the very low energy
contribution to the saturation value of the l.h.s. of the sum rule is marginal,
whereas most of the strength sets in at energies of the order of the meson
mass, where the  neglected  terms of the $K (\bar K)N$ amplitudes are
irrelevant.
Therefore, the  fact that the sum rule is well satisfied when
comparing the l.h.s. with $D_{\bar K}(0,\vec{q}\,)$ indicates that neglecting
the $t$-channel dynamics in the $\bar K N$ interaction is actually quite a good
approximation.  In fact, the omitted mechanisms start appearing
at the one-loop level and involve
the excitation of intermediate $K N$ states in a $t$-channel configuration. 
These $K N$ loop contributions have shown to be relatively weak in the
dynamics of the crossed $s$-channel configuration of the $K N$ system.
Conversely, the neglected 
$t$-channel meson-baryon loop terms in the $K N$ scattering amplitude 
involve the excitation of both $S=-1$, $\bar K N$ and $\pi Y$,
intermediate states, which have been shown to interact quite
strongly in the $s$-channel configuration present in $\bar K N$ dynamics.
It is then clear that the calculated $\bar K$ propagator is more accurate since
the neglected terms are smaller, a fact that is corroborated by the better
fulfillment of the  $m_{-1}$ sum rule in this case.

The $m_0^{(-)}$ sum rule tells us that the areas subtended by the $K$ and $\bar
K$ spectral functions should coincide. This is indeed the case for the
calculation considered here, as can be seen in the middle panel of
Fig.~\ref{fig:kaon-cold}.
We would like to emphasize here that the fulfillment of the $m_0^{(-)}$ sum
rule for the model of kaon interactions under analysis is far from being
trivial. We recall that whereas one expects the $\bar K$ and $K$ spectral
functions to be related by the retardation property, $S_{\bar
K}(-\omega)=-S_K(\omega)$, the actual calculation of the meson self-energies is
done exclusively for positive meson energies (time-like region in the $\bar K
(K) N$ scattering amplitude). 
The analytical constraints are nevertheless imposed in the self-consistent
evaluation of the scattering amplitudes and self-energies \cite{Tolos:2008di}
through the use of
$\bar K$ and $K$ in-medium propagators in the form of
Eq.~(\ref{eq:prop_lehmann}), which couples the information of the two spectral
functions. We note, for instance, that a simplified mean-field like
description of the
meson self-energies by means of effective in-medium masses, namely
$S(\omega,\vec{q}\,)=\delta[\omega-\omega^*(\vec{q}\,)]/2 \omega^*(\vec{q}\,)$
with $\omega^*(\vec{q}\,)=\sqrt{\vec{q}\,^2+m^*{}^2}$, would clearly violate the
$m_0^{(-)}$ sum rule since $\Delta  m_{\bar
K}^*(\rho)  < 0$ and $\Delta m_{K}^*(\rho) > 0$.

The $m_0^{(+)}$ sum rule saturates to one independently of the meson momentum,
nuclear density or temperature, thus posing a strong constraint on the accuracy
of the calculations. It has been thoroughly used to test the quality of the
nucleon spectral function in the nuclear many-body problem.
The lower panel in Fig.~\ref{fig:kaon-cold} shows that the calculated $K$ and
$\bar K$ spectral functions fulfill this sum rule to a high
precision. The particle and anti-particle parts converge to
different values in general, but the sum perfectly saturates to the required
value of one. Also note that saturation is progressively shifted to higher energies as
we examine sum rules involving higher order weights in energy.

\begin{figure}[t]
\begin{center}
\includegraphics[width=0.8\textwidth]{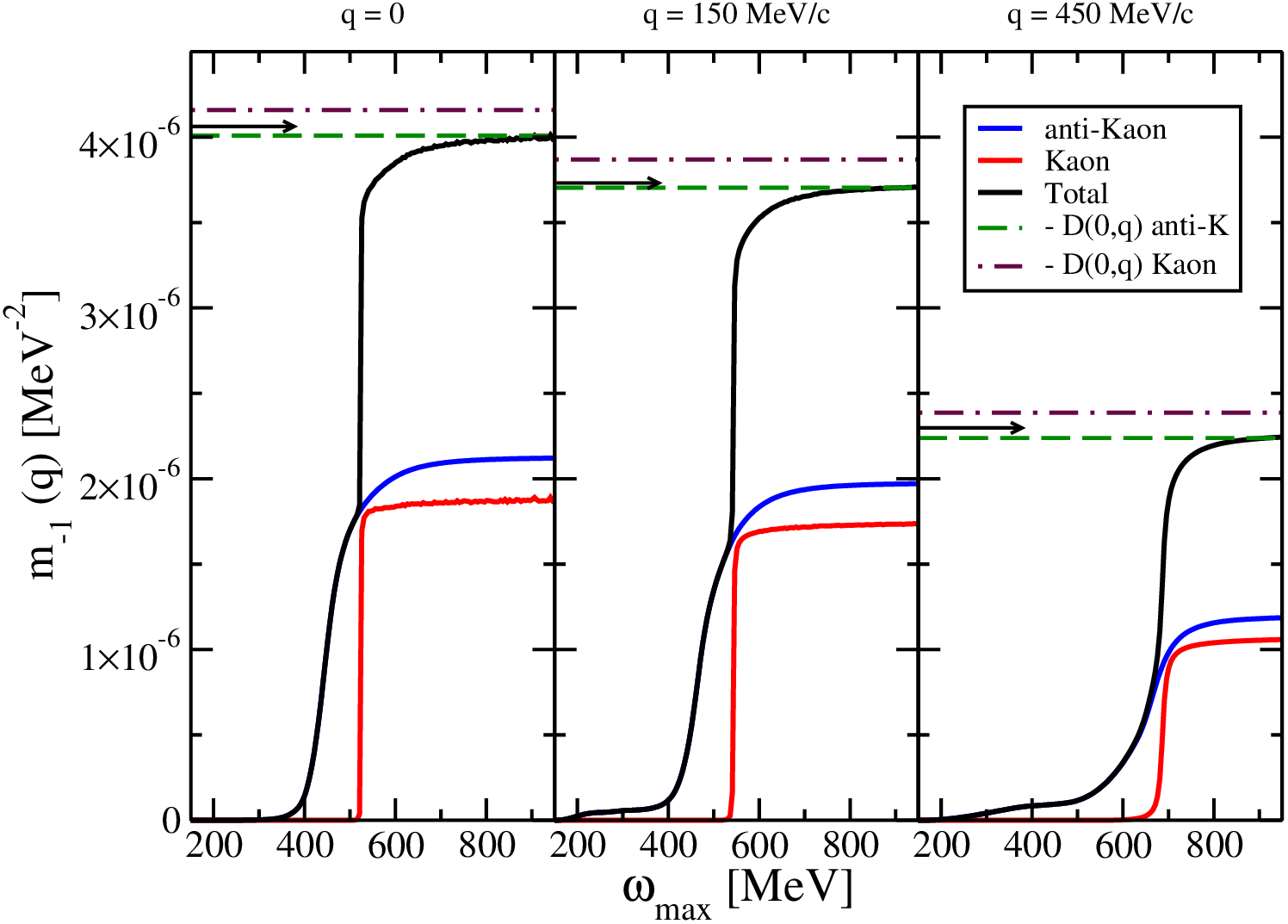}
\\
\includegraphics[width=0.8\textwidth]{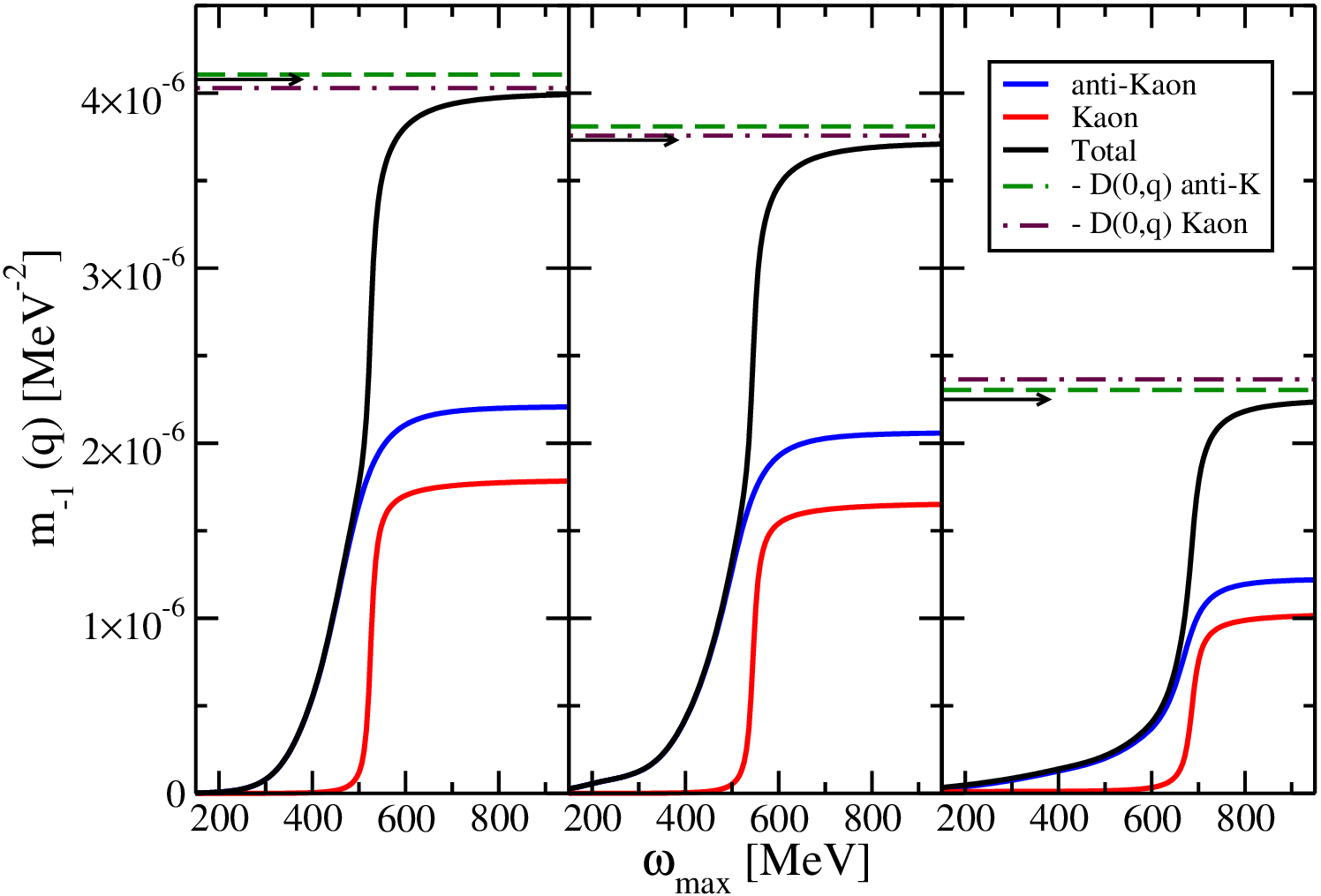}
\caption{(Color online) $m_{-1}$ sum rule for the $K$ and $\bar K$ spectral functions at
several momenta ($q=0,150,450$~MeV/c) and $\rho=\rho_0$. Upper panel: zero temperature. Lower panel:
$T=100$~MeV. The arrows indicate the value of the r.h.s. in vacuum.}
\label{fig:kaon-minus-one-variosq}
\end{center}
\end{figure}

\begin{figure}[t]
\begin{center}
\includegraphics[width=0.8\textwidth]{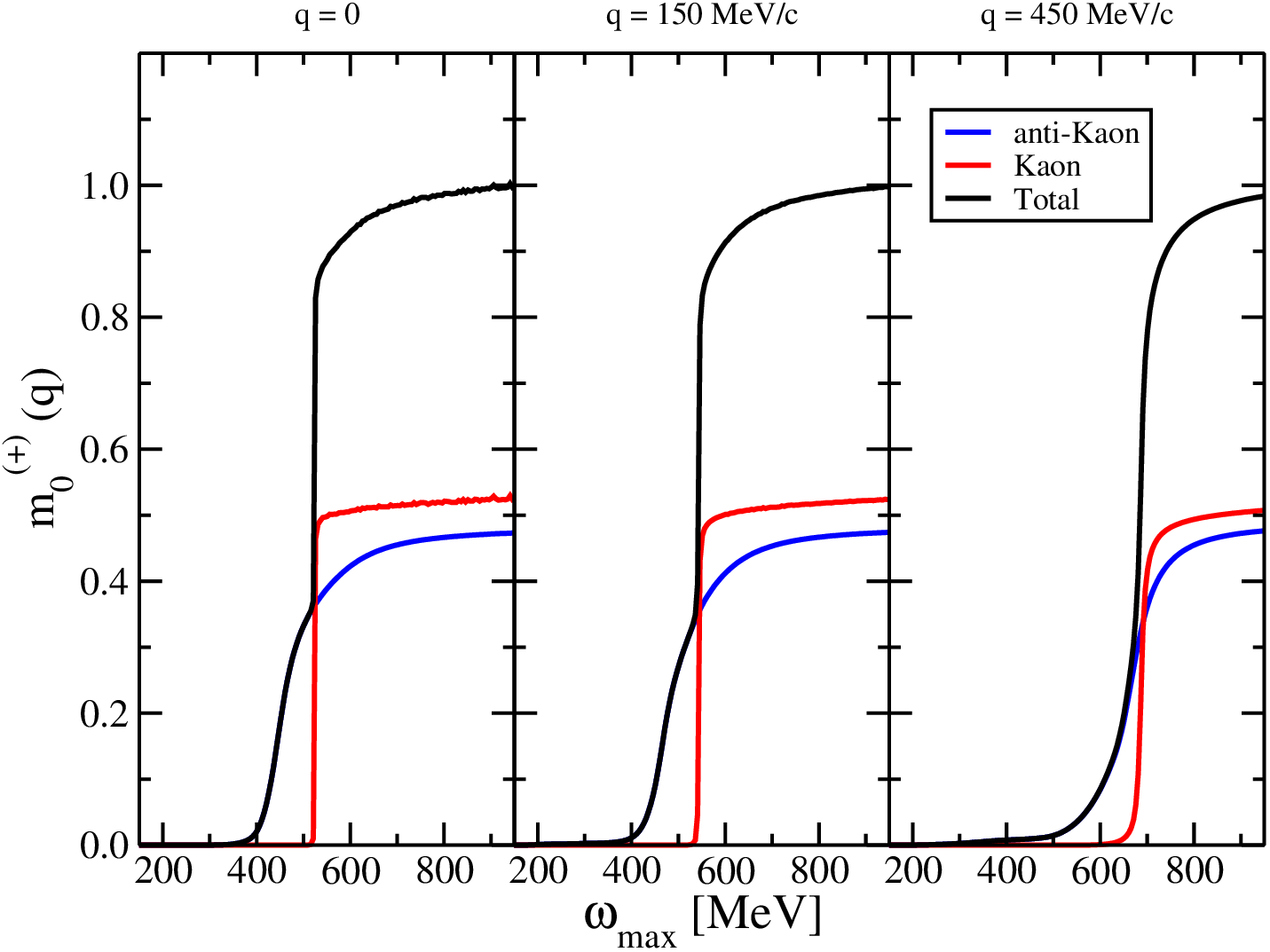}
\\
\includegraphics[width=0.8\textwidth]{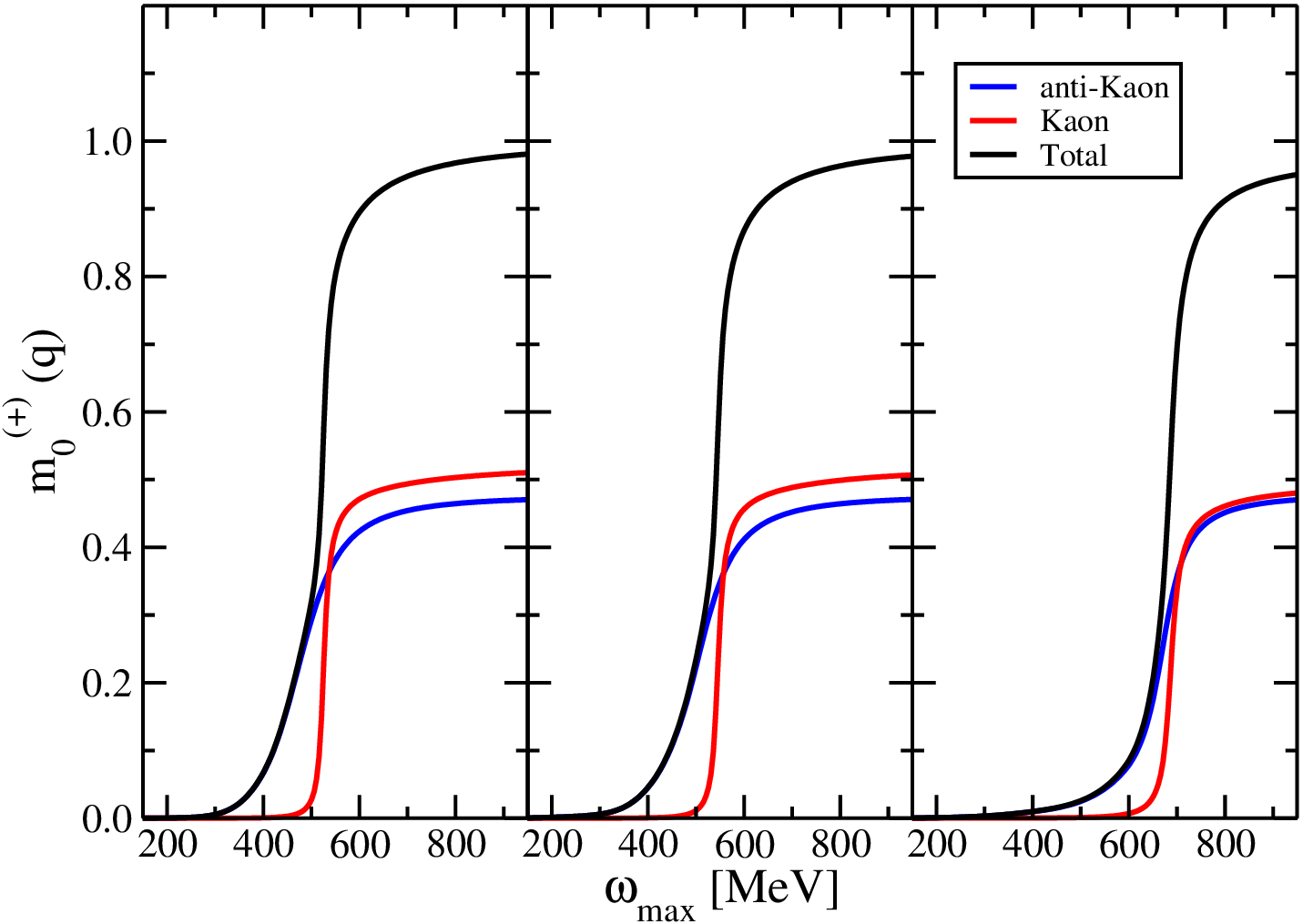}
\caption{(Color online) $m_{0}^{(+)}$ sum rule for the $K$ and $\bar K$ spectral functions at
several momenta ($q=0,150,450$~MeV/c) and $\rho=\rho_0$. 
Upper panels: zero temperature. Lower panels:
$T=100$~MeV.}
\label{fig:kaon-zero-plus-variosq}
\end{center}
\end{figure}

Next we show in Figs.~\ref{fig:kaon-minus-one-variosq} and
\ref{fig:kaon-zero-plus-variosq} the results for $m_{-1}$ and $m_0^{(+)}$,
respectively, at normal nuclear density for different kaon momenta, at zero
temperature (upper panels) and $T=100$~MeV (lower panels). 
As the meson momentum is increased, the saturation
of the integral part of the sum rules is progressively shifted to higher
energies, following the strength of the spectral distribution. In particular,
$m_{-1}$ exhibits a growing sensitivity to the low-energy $P$-wave $Yh$ modes,
which are enhanced at finite momentum. At finite temperature
the $\bar K$ spectral function spreads considerably \cite{Tolos:2008di}, and
in particular acquires a sizable low energy tail from smearing of the Fermi
surface, which contributes substantially to the l.h.s. of the sum rule below
the quasi-particle peak. Note also that the $K$ contribution
softens at finite
temperature and increasing momenta, as the $K$ in-medium decay width is
basically driven by the $KN$ thermal phase space.

We observe that the $m_{-1}$ sum rule is well satisfied by the zero-temperature
$\bar K$ spectral function for the three different momenta represented in the
plot, which reinforces our discussion about the model approximations elaborated
above (we have also checked the momentum dependence of the saturated sum rule in
a wide range of momenta from 0 to 1~GeV/c with the same conclusions). At finite
temperature, however, there is no longer a good agreement for the anti-kaon.
This is not  a failure of the model interaction but of the calculation itself:
the expression in Eq.(\ref{eq:selfd}) for the (dominant) S-wave anti-kaon
self-energy  contains in fact an approximation to the dispersive contribution,
as explained in Appendix B of \cite{Tolos:2008di}, which is appropriate for
energies close to the kaon mass or higher. It  can also be seen there that the
dispersive contributions to the kaon and anti-kaon self-energy practically
vanish at $q^0=0$ (this is exact for the imaginary part of the selfenergy),
while our approximated finite-$T$ expression, Eq.~(\ref{eq:selfd}), does not
satisfy this requirement. Therefore, it is expectable that neither the kaon nor
the anti-kaon self-energies fulfill properly the $m_{-1}$ sum rule at finite T
under the current approximations. Moreover, we observe that the discrepancy of
either contribution with the l.h.s. is of the same order of magnitude as if we
had just used the free propagator to evaluate the r.h.s. As a consequence, one
may not employ this sum-rule as a way to test the interaction model, unless a
very refined finite-$T$ calculation of the self-energy is performed in the far
off-shell, space-like energy region.

The $m_0^{(+)}$ sum rule is
fulfilled satisfactorily for the different momenta and the two temperatures
considered. We note, though, that convergence turns slower for increasing
momentum and finite temperatures and, in some of the cases shown and
up to the maximum energy explored, the
limiting value of one has not yet been reached.

We have checked that, for a wide range of momentum values, the $m_0^{(-)}$ sum
rule at $T=0$ converges to zero with similar precision as the $q=150$ MeV/c case
shown in Fig.~\ref{fig:kaon-cold}. At finite $T$, the agreement is slightly worse
but consistent with zero, admitting a 3\% error in the calculated spectral
functions. In the case of the $m_{1}^{(-)}$ sum rule, the additional $\omega^2$
weight tends to magnify the numerical inaccuracies or model deficiencies of the
spectral functions. Nevertheless, for any of the two temperatures explored in
this work, we find that the finite value at which the $m_{1}^{(-)}$ sum rule
saturates is compatible with zero if one admits a 10\% numerical error in our
calculated spectral function at energies around 1 GeV.
Finally, we do not
evaluate the $m_1^{(+)}$ sum rule for our model calculation of the kaon spectral
functions. The reason is twofold:  on the one hand, the $K(\bar K)$ spectral
function has only been calculated up to about 1~GeV,  due to limitations in the
validity of the chiral unitary amplitudes, while the $m_1^{(+)}$ sum rule
carries a $\omega^3$ energy weight and thus the contribution of higher energies
is still relevant in establishing the saturation of the l.h.s.  On the other
hand, the non-dispersive contribution of the self-energy, associated to the high
energy limit of the interaction and entering the r.h.s., corresponds in our
model to the tree level $\bar K (K)N$ vertex from the meson-baryon chiral
Lagrangian. Its essentially linear energy dependence cannot be extrapolated to
high energies without introducing hadronic form factors (and thus additional
free parameters), leaving the actual value of $\Pi^{\infty}_{\bar K (K)}$
unconstrained.

The pion sum rules are discussed in the following. They exhibit notable
differences with respect to those of the kaon propagator. To start with,
there are no
sum rules weighted with even energy powers, so we shall present results on $m_{-1}$, $m_0$ and $m_{1}$. Second,
the pion spectral function at intermediate momenta displays well separated
collective modes, particularly at very low energies, which can be easily
tracked in the sum rule saturation. The changes introduced by finite
temperature are also visible and worth discussing.
\begin{figure}[t]
\begin{center}
\includegraphics[width=\textwidth]{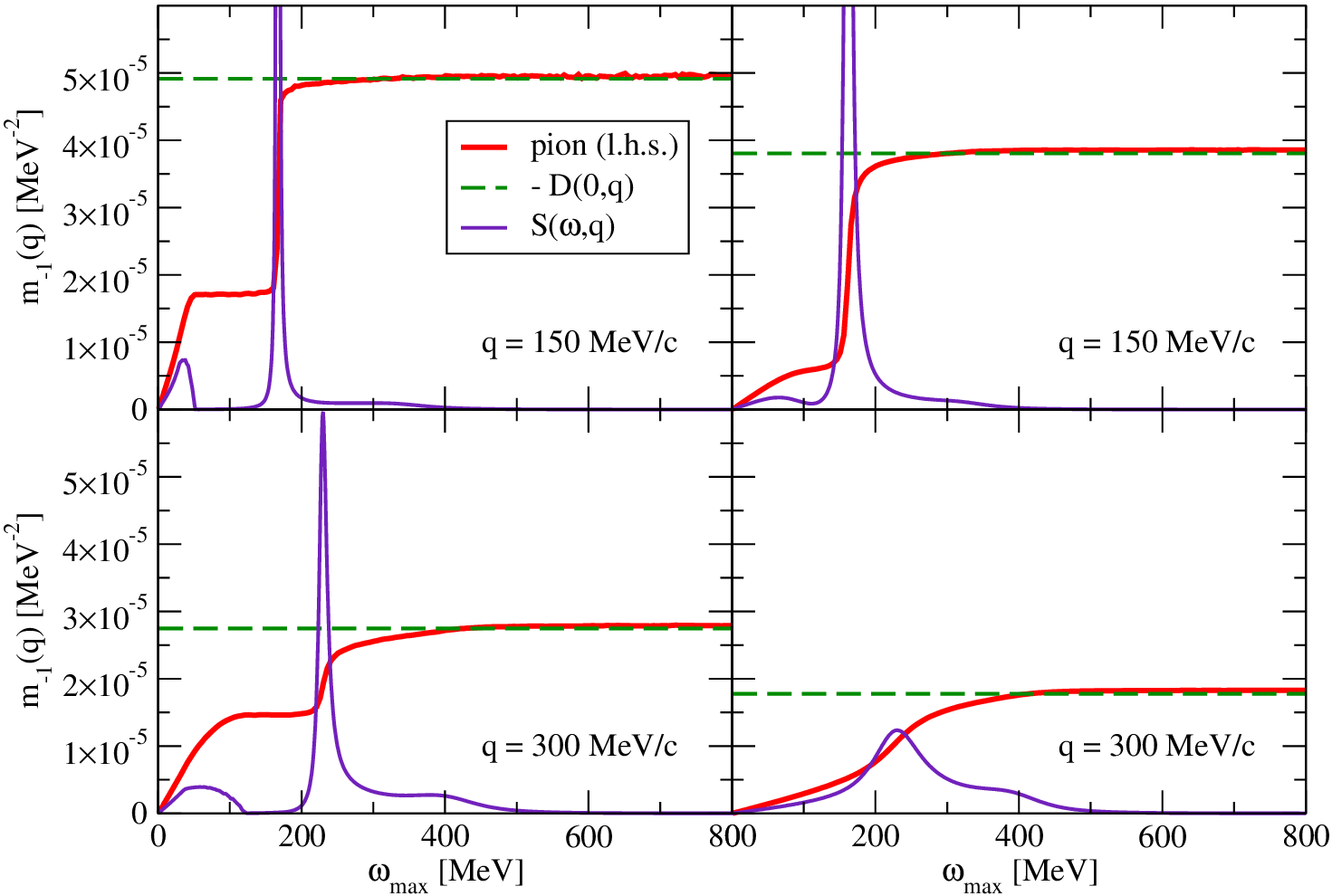}
\caption{(Color online) $m_{-1}$ sum rule for the pion spectral function at
$q=150,300$~MeV/c and $\rho=\rho_0$. The left column corresponds to the zero
temperature result and the right column to $T=100$~MeV. The pion spectral
function is also displayed for reference in arbitrary units.}
\label{fig:pion-minusone-and-zero}
\end{center}
\end{figure}

In Fig.~\ref{fig:pion-minusone-and-zero} (left column) we show $m_{-1}$ for two
pion momenta, at normal nuclear density and
zero temperature. In the case of $q=150$~MeV/c most of the
strength of the spectral function is carried by the pion quasi-particle peak.
However, the $ph$ component lies at low energies and therefore is rather
enhanced in the integral by the inverse energy weight, contributing in more
that 30\% to the saturation value, whereas the pion quasi-particle peak
practically carries the remaining strength. The $\Delta h$ mode is barely
visible to the right-hand side of the pion peak and contributes little at small
momentum. Note the plateau
in $m_{-1}$ due to the energy gap between the low energy part of the spectrum
and the pion mode. At higher momentum ($q=300$~MeV/c), the $ph$ and $\Delta h$
excitations acquire more relevance with respect to the quasi-pion.
The l.h.s. saturates and is in good agreement with the r.h.s., 
slightly overshooting
the value of the inverse pion propagator at zero energy for increasing momentum.
We understand these tiny deviations as originating by the implementation of the
$\Delta$ decay width in the $\Delta h$ contribution, which may lead to small 
violations of the analytical properties of the
pion self-energy and propagator. Essentially, the $\Delta$ self-energy employed
here accounts for the $\Delta$ width and its energy dependence coming from
its decay to a $\pi N$ pair in $P-$wave. However, the
nucleon momentum in the $\Delta h$ excitation is averaged out to make 
the $\Delta$ self-energy dependent only on the pion external four-momentum $q$.
The fulfillment of the $m_{-1}$ sum rule therefore indicates that, 
up to tiny deviations tied to these
kinematical averages, accounting for the $\Delta$ decay width and its energy dependence 
not only provides a more
realistic description of the phenomenology, but additionally complies 
with analyticity requirements
through the sum rules. As a test, we have also calculated $m_{-1}$ with a 
constant $\Delta$ decay width
and the agreement between l.h.s. and
r.h.s. of the sum rule is far worse than in the present model.

At finite temperature, the softening of
the nucleon occupation number due to thermal motion causes the broadening of
the three modes, as can be seen in the right column of
Fig.~\ref{fig:pion-minusone-and-zero}.
At a momentum value of 300 MeV/c they are rather mixed which removes the
plateau visible at zero temperature (left panels) and the contribution from
each excitation mechanism to the l.h.s. of the sum rule can no longer be 
resolved.
\begin{figure}[t]
\begin{center}
\includegraphics[width=0.8\textwidth]{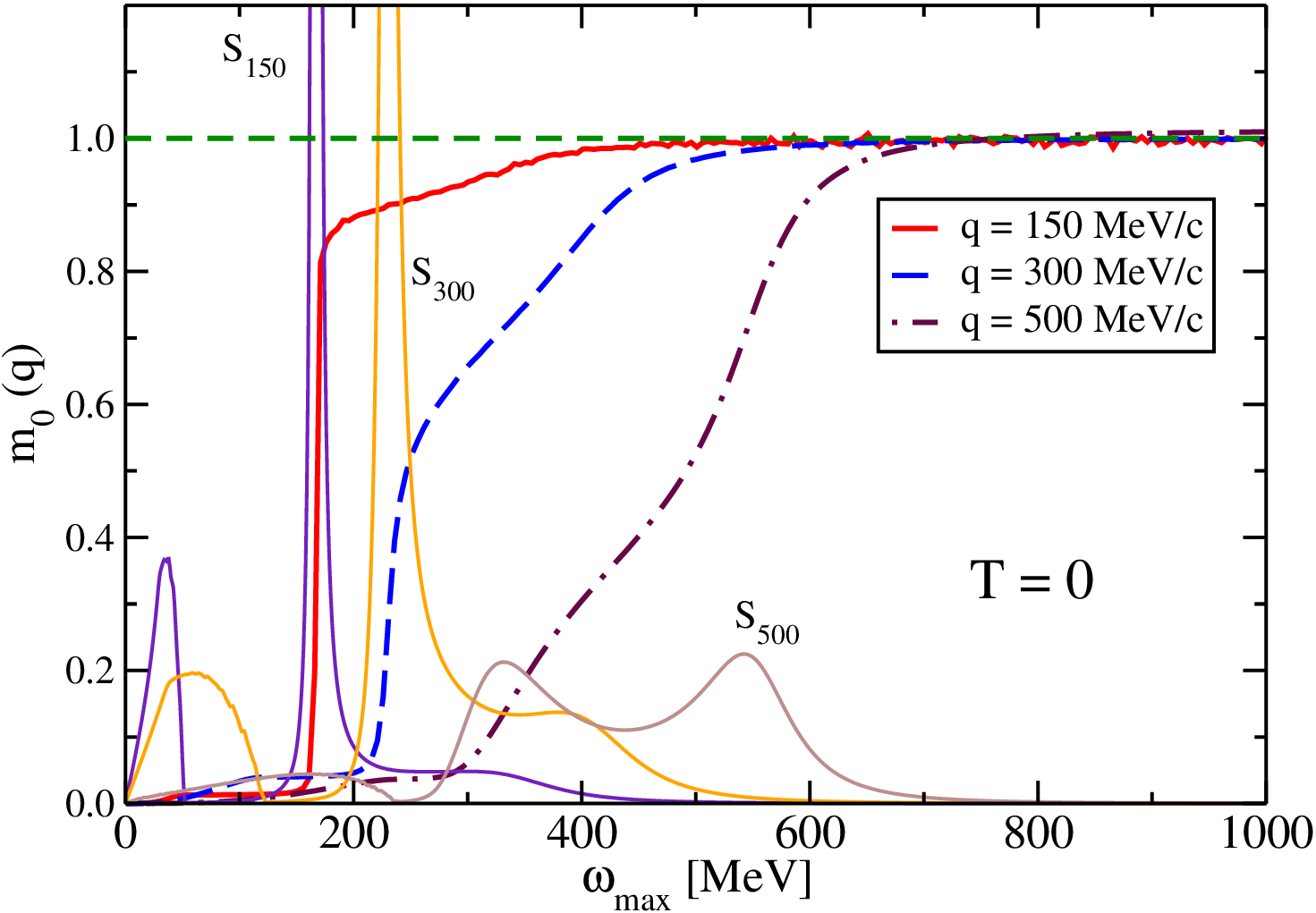}
\\
\includegraphics[width=0.8\textwidth]{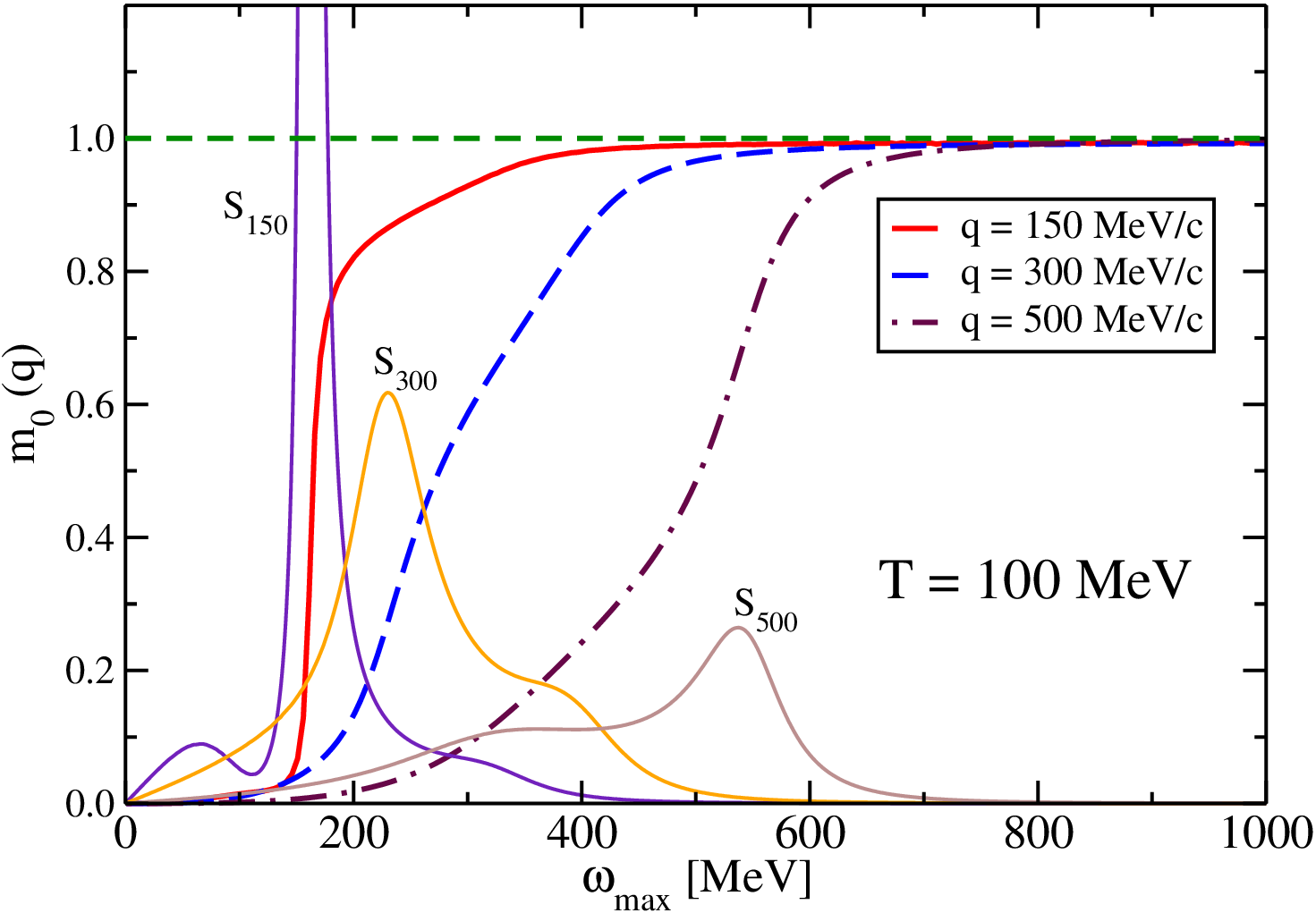}
\caption{(Color online) $m_{0}$ sum rule for the pion spectral function at several momenta,
$\rho=\rho_0$ and $T=0,100$~MeV. The pion spectral functions are also displayed for reference in
arbitrary units, labelled as $S_q$ with $q$ the corresponding momentum in MeV/c. Note that this sum
rule is independent of the meson momentum.}
\label{fig:pion-zero-various-q-and-T}
\end{center}
\end{figure}

The $m_0$ sum rule is depicted in Fig.~\ref{fig:pion-zero-various-q-and-T}
for three values of the pion momenta and $T=0,100$~MeV. Despite the markedly
different distribution of strength in the spectral density with increasing
momentum and temperature, the sum rule is quite well satisfied in all
cases. The deviations observed may also be attributed to the small violations
of the analytical properties of the pion self-energy mentioned above.

\begin{figure}[t]
\begin{center}
\includegraphics[width=0.8\textwidth]{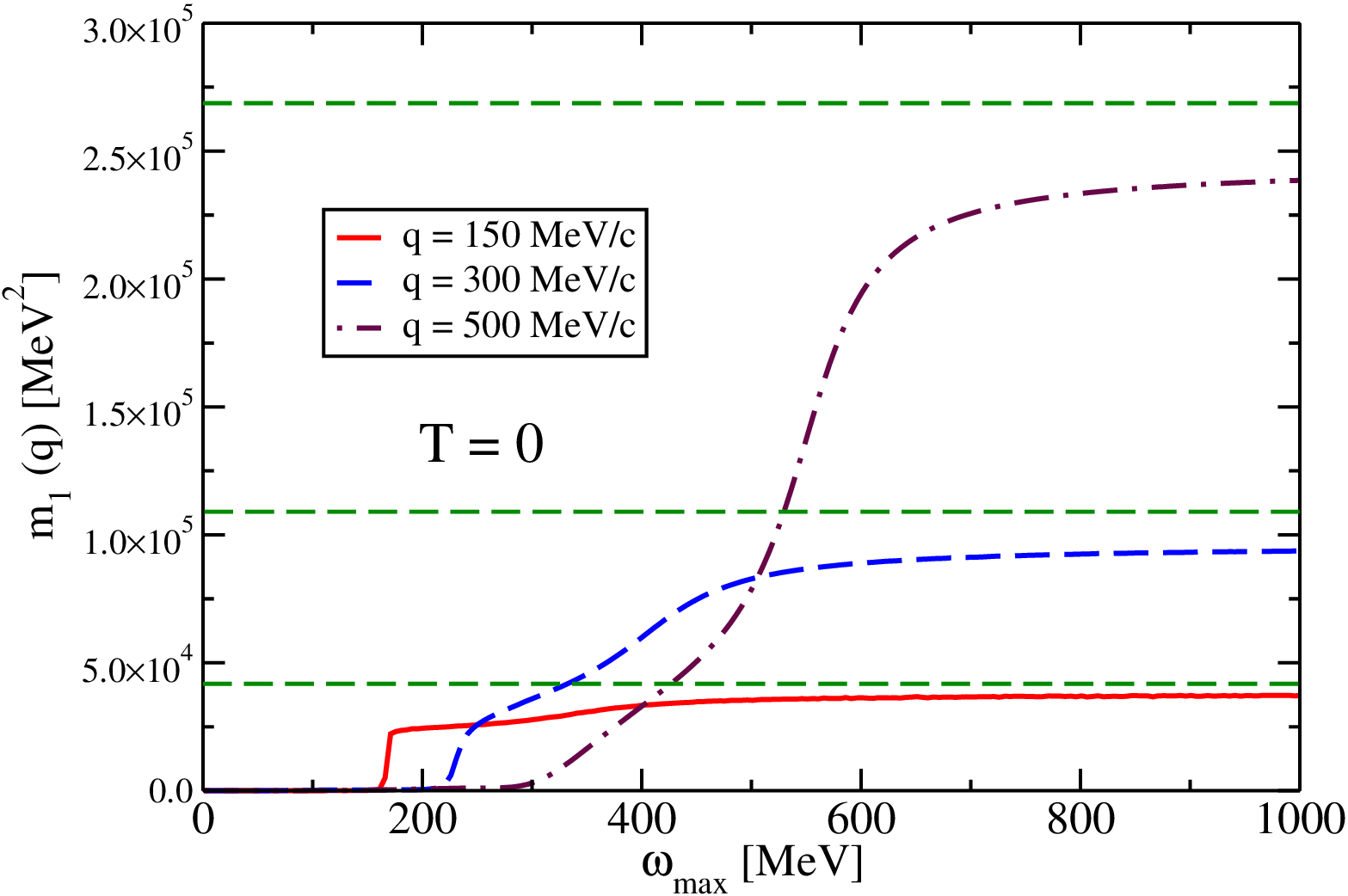}
\\
\includegraphics[width=0.8\textwidth]{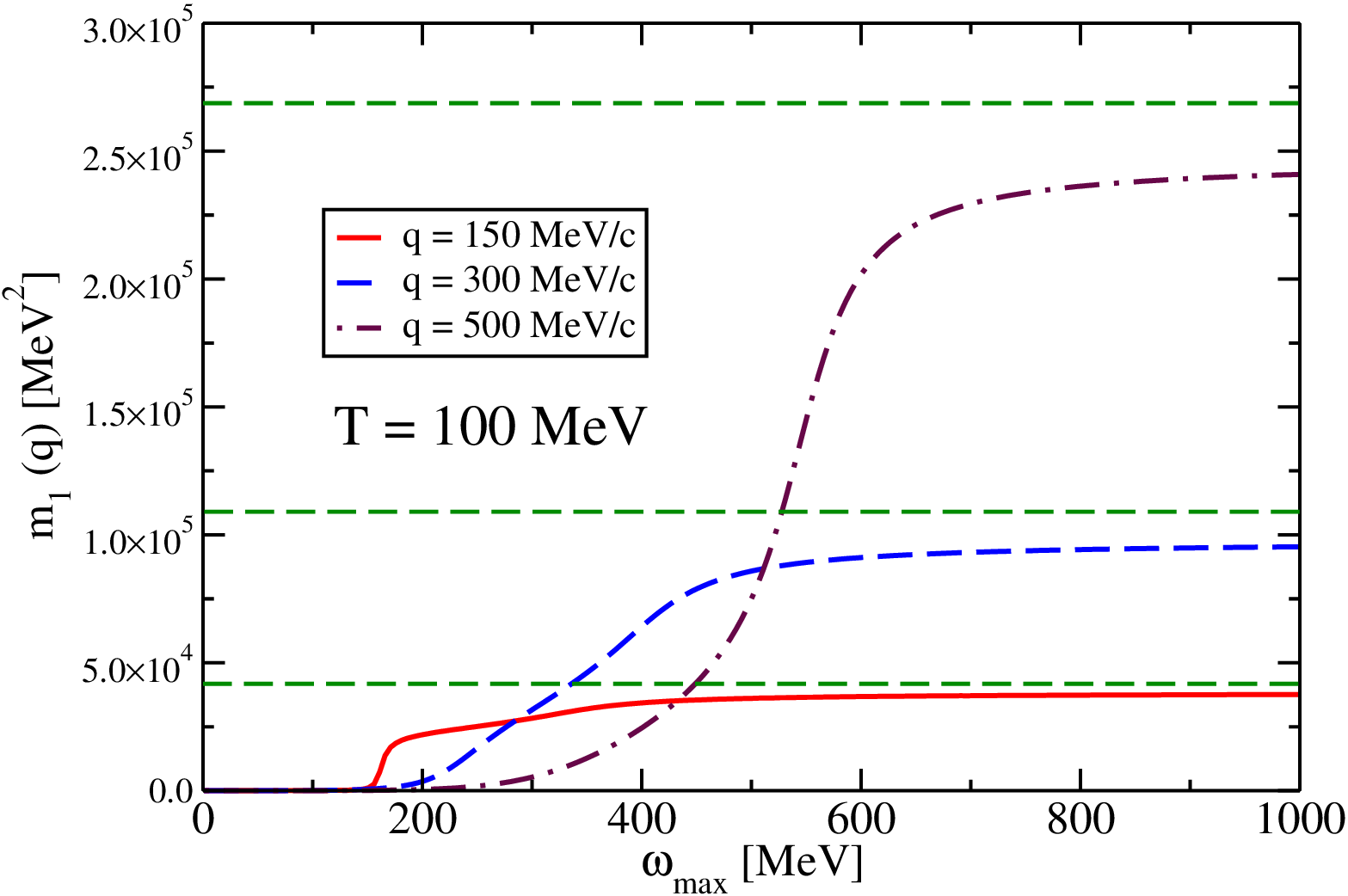}
\caption{(Color online) $m_{1}$ sum rule for the pion spectral function at several momenta,
$\rho=\rho_0$ and $T=0,100$~MeV. The dashed, horizontal lines correspond to the
r.h.s. of the sum rule for each momentum,
$\omega_{\pi}^2=m_{\pi}^2+\vec{q}\,^2$.}
\label{fig:pion-one-various-q}
\end{center}
\end{figure}

Finally we show $m_1$ for the pion propagator in
Fig.~\ref{fig:pion-one-various-q}, for several momenta and two temperatures,
$T=0$ and 100~MeV. The $m_1$ sum rule carries a $\omega^3$ energy weight which
makes it sensitive to higher energies and thus its convergence is much slower.
Conversely, the contribution from low energy modes is marginal in this case. The
pion self-energy analyzed here, built from the tree-level $p$-wave coupling to
$ph$ and $\Delta h$ modes, admits a purely dispersive representation, as it can
be easily derived from the Lindhard function. Therefore,  $\Pi^{\infty}_{\pi}=0$
in our model and the l.h.s. of the sum rule is seen to slowly converge to the
squared single-particle pion energy in vacuum,
$\omega_{\pi}^2=m_{\pi}^2+\vec{q}\,^2$.

\section{Summary and outlook}
\label{sec:conclusions}

In summary, we have presented a derivation of the energy weighted sum rules of
the meson propagator in nuclear matter, which can be applied to a wide range of
scenarios such as meson systems with a different in-medium behavior of
particle and anti-particle modes, isospin-asymmetric matter, and matter at 
finite temperature. We have particularized the sum rules for kaons and
pions in cold and hot symmetric nuclear matter, where specific models for the meson
self-energy and spectral function have been analyzed from the point of view of
the saturation of the sum rules.

The sum rules have been shown to be a useful tool to magnify troublesome
situations where certain approximations typically done in the calculation of
the scattering amplitudes (and thus, of the meson self-energies) may not work
properly in particular kinematical regimes. This is possible since the sum
rules explored relate the energy-weighted meson spectral function, integrated
over all energies, to the meson propagator, evaluated at low or high energies,
as well as to model independent quantities. For instance, thanks to visible deviations
in the $m_{-1}$ sum rule, we have seen that
violation of crossing symmetry in the chiral
unitary $K(\bar K)N$ interaction model employed becomes relevant
at time-like energies. The $m_{-1}$ sum rule is 
not properly fulfilled in the case of pions either
 if a constant (energy-independent) 
 $\Delta$ decay width is employed in the intermediate $\Delta h$ excitations.
An oversimplified description of medium
effects on the meson single particle properties, such as the use of
effective in-medium masses or approximating meson spectral functions by
quasiparticle Breit-Wigner peaks ---thus ignoring the role of $N h$, $\Delta h$,
$Y h$ or $Y^* h$ components--- may lead to
violations of the sum rules already at the lowest orders. 
We also note that, even if the interaction model fulfilled the proper
analyticity requirements,
certain sum rules may also be useful to check the accuracy of the numerical
evaluation of the meson spectral functions, as is the case of the momentum 
independent $m_0$.

The present work can be used to study the quality of many-body approaches and
interaction models in other systems such as light vector and axial-vector meson
resonances, where a straightforward extension of the formalism is required in
order to describe transverse and longitudinal modes. The study of open- and
hidden-charm meson resonances in hot and dense matter has also received much
interest lately and will be subject of experimental investigation in heavy-ion
experiments at FAIR. Present and future calculations of the interaction of
these systems with the medium can also be scrutinized from the point of view of
EWSR's. Work along these lines is in progress.

\section*{Acknowledgments}
This work is partly supported by the EU contract No. MRTN-CT-2006-035482
(FLAVIAnet), by the contracts FIS2008-01661/FIS and FPA2008-00592 from MICINN
(Spain) and by the Ge\-ne\-ra\-li\-tat de Catalunya contract 2005SGR-00343. We
acknowledge the support of the European Community-Research Infrastructure
Integrating Activity ``Study of Strongly Interacting Matter'' (HadronPhysics2,
Grant Agreement n. 227431) under the Seventh Framework Programme of EU, the
``RFF-Open and hidden charm at PANDA'' project from the Rosalind Franklin
Programme of the University of Groningen and the Helmholtz International Center
for FAIR within the framework of the LOEWE program (Landesoffensive zur
Entwicklung Wissenschaftlich-Oekonomischer Exzellenz) launched by the State of
Hesse (Germany). D.~Cabrera acknowledges support from the "Juan de la Cierva"
Programme (MICINN, Spain).


\end{document}